\documentclass[12pt]{article}

\usepackage{amsmath,amsfonts}
\usepackage{amssymb}
\usepackage{epsfig,psfrag,cite}
\usepackage[usenames,dvips]{color}

\usepackage{multirow}

\usepackage{stackrel}
\usepackage{subfigure}
\usepackage[margin=3.5cm]{geometry}

\makeatletter
\@addtoreset{equation}{section}
\makeatother

\newcommand{\be}{\begin{equation}}
\newcommand{\ee}{\end{equation}}
\newcommand{\bea}{\begin{eqnarray}}
\newcommand{\eea}{\end{eqnarray}}

\newcommand{\nn}{\nonumber}

\begin{document}
\thispagestyle{empty}

\begin{center}
\hfill CERN-PH-TH/2009-141\\
\hfill UAB-FT-670

\begin{center}

\vspace{1.7cm}

{\LARGE\sc Soft-wall stabilization}

\end{center}

\vspace{1.4cm}

{\bf Joan A. Cabrer$^{\,a}$, Gero von Gersdorff$^{\,b}$ and Mariano Quir\'os$^{\,a,\,c}$}\\

\vspace{1.2cm}
${}^a\!\!$ {\em {IFAE, Universitat Aut{\`o}noma de Barcelona,
08193 Bellaterra, Barcelona, Spain}}

${}^b\!\!$ {\em {CERN Theory Division, CH-1211 Geneva 23, Switzerland}}

${}^c\!\!$ {\em {Instituci\'o Catalana de Recerca i Estudis  Avan\c{c}ats (ICREA)}}

\end{center}

\vspace{0.8cm}

\centerline{\bf Abstract}
\vspace{2 mm}
\begin{quote}\small
We propose a general class of five-dimensional soft-wall models with AdS metric near the ultraviolet brane and four-dimensional Poincar\'e invariance, where the infrared scale is determined dynamically. A large UV/IR hierarchy can be generated without any fine-tuning, thus solving the electroweak/Planck scale hierarchy problem. 
Generically, the spectrum of fluctuations is discrete with a level spacing (mass gap) provided by the inverse length of the wall, similar to RS1 models with Standard Model fields propagating in the bulk. Moreover two particularly interesting cases arise. They can describe: (a) a theory with a continuous spectrum above the mass gap which can model unparticles corresponding to operators of a CFT where the conformal symmetry is broken by a mass gap, and; (b) a theory with a discrete spectrum provided by linear Regge trajectories as in AdS/QCD models.    

\end{quote}

\vfill

\section{Introduction}
Warped extra dimensions were introduced by Randall and Sundrum~\cite{RS1} (RS1) as a very elegant way of solving the hierarchy problem by means of the geometry of extra-dimensional theories. The original proposal consisted of a slice of AdS space bounded by two branes: the ultraviolet (UV) brane, located closely to the AdS boundary, and the infrared (IR) one. One of the most exciting aspects of RS1 theories is provided by the AdS/CFT correspondence~\cite{AdS/CFT}, by which fields living in the UV brane are fundamental fields which can interact with a strongly coupled conformal sector (CFT), while fields living in the IR brane are the dual description of operators of the CFT. In this way a Higgs boson localized on the IR brane provides a dual description of the composite Higgs as a bound state of an extra strong interaction (technicolor). From the five dimensional (5D) point of view the Higgs boson mass is simply redshifted from its natural value, of the order of the Planck mass, to the TeV scale by the warp factor. The IR brane spontaneously breaks the conformal symmetry and its location needs to be stabilized by some dynamical mechanism.

A dynamical mechanism to stabilize the IR brane was proposed by Goldberger and Wise~\cite{GW} (GW) by the introduction of a background scalar field propagating in the 5D bulk and acquiring a coordinate dependent vacuum expectation value (VEV). This triggered a 4D effective potential for the radion field with a minimum, stabilizing the brane separation at scales of order 1/TeV, determined by the values of the scalar field at both the UV and IR branes, with a modest $\sim 1\%$ fine-tuning. Furthermore the back-reaction of the scalar field through the 5D Einstein equations generated a deviation from AdS of the metric far away from the UV brane, still preserving the main features of the AdS/CFT correspondence. In fact this is a general feature since AdS is the only 5D metric consistent with no (or constant) background scalar field, which means that any stabilizing bulk field is expected to back-react on the AdS metric. 

Since the presence of bulk scalar fields, required to stabilize the brane distance, is expected to disturb the AdS metric in the far IR, one can introduce phenomenological models with background scalar fields and with the only constraint of describing the AdS geometry near the UV brane. One of them is of course RS1 with GW mechanism, but there are more general models~\footnote{For an earlier analysis of models with exponential (Liouville) dilatonic brane potentials see Ref.~\cite{Chamblin:1999ya}.} where the IR brane is replaced by a naked curvature singularity~\cite{ArkaniHamed:2000eg,Kachru:2000hf,Gubser:2000nd,Csaki:2000wz}: these models are called {\it soft-wall} models~\cite{AdS/QCD,Gursoy:2007cb ,Batell1,Falkowski1,Batell2,Falkowski:2008yr}. Soft-wall models possess AdS geometry near the UV brane and no IR brane, i.e.~they have a non-compact extra dimension as those proposed by Randall and Sundrum with an infinite extra dimensional length~\cite{RS2} (RS2), but with additional background scalar fields. These fields back-react on the metric and generate a singularity at a finite value of the extra dimensions: the (finite) length of the extra dimension provides the location of the soft wall.   

Soft wall models were first introduced to model the Regge behaviour of excited mesons in AdS/QCD models~\cite{AdS/QCD,Gursoy:2007cb}, as an alternative to RS1 for electroweak breaking models solving the hierarchy problem and providing experimental signatures at LHC~\cite{Batell2,Falkowski1} and also to provide a 5D setup to the theories of unparticles recently proposed by Georgi~\cite{Georgi:2007ek} in the presence of a mass gap~\cite{Cacciapaglia:2008ns,Falkowski:2008yr}. In all these theories, as in GW models, there are background bulk scalar fields and thus a built-in mechanism to stabilize the extra dimension, i.e.~the distance between the UV brane and the singularity. However in the proposed models this distance is naturally of the order of the AdS length and the hierarchy problem does not find a natural solution, i.e.~for values of the scalar field at the UV brane of the order of the 5D Planck scale~\cite{Batell2}.

In this paper~\cite{Planck2009} we will propose a set of one-parameter ($\nu$) 5D models with a background scalar field propagating in the bulk of the extra dimension and with the following properties:
\begin{itemize}
\item
The metric is AdS near the UV brane.
\item
 The hierarchy between the AdS length and the soft-wall length (the electroweak length) is naturally stabilized for values of the scalar field at the UV brane of the order of the 5D Planck scale. The naturalness of this result comes from a double exponential suppression. 
\item
For values of $\nu<2$ the naked singularity does not contribute to the vacuum energy and thus it does not need to be resolved to satisfy Einstein equations. Moreover the potential is bounded from above in the solution and the singularity is a physical one according to Ref.~\cite{Gubser:2000nd}.
\item
For $1<\nu<2$ the spectrum of fluctuations behave as in RS1 models with Standard Model (SM) fields propagating in the bulk: for scalar field fluctuations localized near the singularity the hierarchy problem is automatically solved.
\item
For $\nu=1$ one can model unparticles with a mass gap provided by the inverse length of the extra dimension. Fluctuations have a continuous spectrum above the mass gap.
\item
For $0<\nu<1 $ unparticles without a mass gap emerge: the spectrum is continuous above zero mass.
\end{itemize}

The paper is organized as follows: In Section~\ref{general} we review the general construction of backgrounds and the associated question of self vs.~fine-tuning of the cosmological constant. The general conditions for models with physical singularities are summarized. In Section~\ref{model} we consider the background solutions for a particularly simple class of models satisfying all physical requirements on the singularity and describing the electroweak/Planck hierarchy without fine-tuned parameters. Fluctuations on the background for the graviton, radion and the scalar field are studied in Section~\ref{fluctuations} where a complete numerical analysis and some analytical approximations are provided. A general class of models having all the good required properties, included the hierarchy determination, is presented in Section~\ref{others}. One of these models has the mass of excitations as $m_n^2\sim n$ and it is thus a good candidate to model the Regge behaviour in AdS/QCD models. Finally our conclusions and outlook are drawn in Section~\ref{conclusions}.

\section{The 5D Scalar Gravity System}
\label{general}

In this section we would like to review the construction of
backgrounds with 4D Poincare invariance proposed in
Ref.~\cite{Brandhuber:1999hb ,DeWolfe:1999cp}, and the associated question of self
vs.~fine-tuning of the cosmological constant (CC)~\cite{ArkaniHamed:2000eg,Kachru:2000hf,Gubser:2000nd,Forste:2000ps,Csaki:2000wz,Forste:2000ft}.
Let us consider 5D gravity with a scalar field, and look for the most
general solutions to this system that preserve 4D Poincare invariance,
i.e.~a background of the form
\be
ds^2=e^{-2A(y)}dx^\mu dx^\nu \eta_{\mu\nu}+dy^2\,,
\label{flat}
\ee
with a ``mostly plus'' flat Mikowskian metric $\eta_{\mu\nu} = \operatorname{diag}(-,+,+,+,+)$ and an
arbitrary warp factor $A(y)$.  We will introduce a single brane
sitting at $y=0$ and impose the orbifold $\mathbb Z_2$ symmetry $y\to -y$
under which $A$ and $\phi$ are even.
We will thus consider the bulk plus brane action
\be
S=\int d^5x \, \sqrt{-g}\left[M^3 R-3(\partial \phi)^2-V(\phi)\right]
-\int d^4 x \, \sqrt{-g_{ind}}\lambda(\phi) .
\label{action}
\ee
We have introduced arbitrary bulk and brane potentials $V$ and
$\lambda$, and $M$ denotes the 5D Planck mass which we will set to
unity for the remainder of the paper. Note the noncanonical form of
the kinetic term for the scalar that will simplify future formulas.
The bulk equations of motion (EOM) that follow from the action in
Eq.~(\ref{action}) are
\bea
6\phi''-24A'\phi'-\partial_\phi V(\phi)  &=&0\,,\label{eom1}\\
A''-\phi'^2  &=&0\label{eom2}\,,\\
12A'^2-3\phi'^2+V  &=&0\,.\label{eom3}
\eea
This system has three integration constants~\footnote{Differentiating
the third equation one obtains a linear combination of the first two,
such that (say) the first can be discarded. The system is then first
order in $\phi$ and second order in $A$.}. One of them is $A(0)$ that
remains totally free. The other two can be fixed from the boundary
conditions (BC) that follow from the boundary pieces of the EOM,
\bea
A'(0_+)&=&\frac{1}{6}\lambda(\phi_0)\,,\label{bc1}\\
\phi'(0_+)&=&\frac{1}{6}\partial_\phi\lambda(\phi_0)\,,
\label{bc2}
\eea
where $\phi_0=\phi(0)$.
Using Eqs.~(\ref{bc1}) and (\ref{bc2}) in Eq.~(\ref{eom3}) determines $\phi_0$,
\be
\frac{1}{12}[\partial_\phi \lambda(\phi_0)]^2-\frac{1}{3}\lambda(\phi_0)^2=V(\phi_0)\,,
\ee
which could be used to replace (\ref{bc2}). 

The authors of Ref.~\cite{DeWolfe:1999cp} introduced the following
``trick'' to obtain solutions to this system by defining the so-called
``superpotential'' via the differential equation
\be
3(\partial_\phi W)^2-12 W^2=V\,,
\label{W}
\ee
and writing
\bea
A'&=&W(\phi)\,,\label{1}\\
\phi'&=&\partial_\phi W(\phi)\,,
\label{2}
\eea
while the boundary conditions are satisfied if
\be
W(\phi_0)=\frac{1}{6}\lambda(\phi_0)\,,\qquad 
\partial_\phi W(\phi_0)=\frac{1}{6}\partial_\phi\lambda(\phi_0)
.
\label{W0}
\ee
Again, the system of Eqs.~(\ref{W})--(\ref{2}) has three integration
constants and in principle every solution to Eqs.~(\ref{eom1})--(\ref{eom3}) can be constructed in this way~\cite{DeWolfe:1999cp}.  One
integration constant is the trivial additive constant $A(0)$ that does
not enter in Eq.~(\ref{W0}).  We are left with the integration
constant in Eq.~(\ref{W}) and the value $\phi_0$ to fix the two
constraints Eq.~(\ref{W0}).  The equation for $W$ is a complicated
nonlinear differential equation, and in practice it is often easier to
start with a particular superpotential satisfying the boundary
conditions and deduce the potential needed to reproduce it.

A peculiarity of the scalar-gravity system with one brane is the
appearance of naked curvature singularities at finite proper distance.
In particular, it can easily be checked that if the superpotential $W$
grows faster than $\phi^2$ at large $\phi$, the profile $\phi(y)$
diverges at finite value of $y\equiv y_s$. Moreover the 5D curvature
scalar along the fifth dimension can be written as
\be
R(y)=8(\partial_\phi W(\phi[y]))^2-20 W(\phi[y])^2
,
\ee
so that the curvature in general diverges at $y=y_s$. The
interpretation is that spacetime ends at $y_s$.

Having three integration constants but only two constraints, it seems
that one can obtain flat 4D solutions with fairly generic brane and
bulk potentials without fine-tuning. This miraculous self-tuning
property of the scalar-gravity system was first pointed out in~\cite{ArkaniHamed:2000eg,Kachru:2000hf} and it was further scrutinized in
several papers~\cite{Gubser:2000nd,Forste:2000ps,Csaki:2000wz,Forste:2000ft}. In
particular, the authors of Refs.~\cite{Forste:2000ps,Forste:2000ft}
pointed out that the on-shell Lagrangian, integrated over the fifth
dimension can be written as
\be
\mathcal L^{\rm on-shell}=\frac{1}{3}\lambda(\phi_0)
+\frac{2}{3}\int dy\ e^{-4A} V(\phi[y])  
\ee
(we have set $A(0)=0$). They then make particular choices for $V$ and
$\lambda$ and show that the result is non-vanishing. The
interpretation of this apparent contradiction to the existence of a
flat background is simple: having dynamically generated a new boundary
at the singularity, we must ensure that the boundary pieces of the
equations of motion vanish at $y=y_s$. If this is not the case, the
resulting ``solution'' does in fact not extremize the action,
resulting in a nonzero 4D CC.  This can actually be seen in rather
general terms. Making use of the equations for the superpotential, we
write
\be
e^{-4A(y)}V(\phi[y])= 3\frac{d}{d y}\biggl\{e^{-4 A(y)}W(\phi[y])\biggr\} 
,
\ee
leading to
\be
\mathcal L^{\rm on-shell}=\frac{1}{3}\lambda(\phi_0)-2W(\phi_0)
+2\lim_{y\to y_s} e^{-4A(y)}W(\phi[y])\,. 
\label{onshell}
\ee
The first two terms cancel if  Eq.~(\ref{W0}) is satisfied, while
the last one depends on the particular form of the superpotential.  In order for the last term to vanish,
$W$ needs to grow more slowly than $e^{2\phi}$ at large $\phi$: this
can be seen by using the field $\phi$ itself as a coordinate. The
position of the singularity moves to $\phi(y_s)=\infty$ and the
equation for $A(\phi)$ becomes $A'=W/W'$. It follows that the last
term in Eq.~(\ref{onshell}) goes to a constant for $W\sim e^{2\phi}$,
while it goes to infinity (zero) when $W$ grows faster (slower) than
$e^{2\phi}$. We thus arrive at a simple criterion for the existence of singular solutions:
\be
\parbox{11 cm}{\it A singularity  with  $\phi(y_s)\to\infty$
is allowed if and only if $W(\phi)$  grows asymptotically more
slowly than $ e^{2\phi}$. 
}\label{criterion}
\ee
Notice that a potential ($V$) growing more slowly than $e^{4\phi}$ is only necessary but not sufficient for the validity of (\ref{criterion}), the trivial counterexample being $V\equiv 0$ which has the general solution $W=c\, e^{2\phi}$.
It is instructive to compare our criterion with the one found in Ref.~\cite{Gubser:2000nd} where AdS-CFT duality was used to classify physical singularities. According to Ref.~\cite{Gubser:2000nd} admissible singularities are those whose potential is bounded above in the solution. Inspection of Eq.~(\ref{W}) shows that singularities fulfilling (\ref{criterion}) have a potential that goes to $-\infty$, while those that fail (\ref{criterion}) go to $+\infty$. Although we here employ a much more basic condition (a consistent solution to the Einstein equations), which in particular can be applied to theories without any field theory dual, it is good to know that 
our allowed solutions have potentially consistent interpretations as 4D gauge theories at finite temperature. 

However achieving a superpotential that grows more slowly than
$e^{2\phi}$ requires a hidden fine-tuning of the cosmological constant. 
To see this it suffices to consider a potential that behaves asymptotically as
\be
V\sim  b e^{2\nu\phi}\,.
\ee
Writing $W$ as
\be
W(\phi)=w(\phi) e^{\nu\phi}\,,
\ee
we can express the solutions for $w$ as the roots of
\be
e^{(4-\nu^2)(\phi-c)}=
(2 w+\sqrt{b+4 w^2})^{\pm 2}(\nu w\mp\sqrt{b+4 w^2})^\nu\,,
\ee
where $c$ is an integration constant. For $\nu>2$ this implies that
$w$ asymptotes to a constant
\be
w\sim \pm \sqrt{\frac{b}{\nu^2-4}}\,
\label{wconst}
\ee
at large $\phi$ and for $b>0$.  However for $0<\nu<2$, $w$
generically behaves as
\be
w\sim e^{(2-\nu)\phi}\,.
\ee
Only if we adjust $c\to\infty$ we can achieve that $w$ behaves as in
Eq.~(\ref{wconst}).  In this case, $b$ has to be negative in order to
have a real solution for $W$.

The generic solution to Eq.~(\ref{W}) thus grows as $W\sim
e^{\nu\phi}$ for $\nu\geq 2$ and $W\sim e^{2\phi}$ for $\nu\leq
2$. However, it is possible to arrange for $W\sim e^{\nu\phi}$ in the
latter case by picking a particular value for the integration constant
in Eq.~(\ref{W})~\footnote{Similar reasonings apply to potentials
that grow even more slowly, e.g., as a power. The generic solution
behaves as $e^{2\phi}$, but particular solutions may exist that behave
as $\sqrt{V}$ and hence allow for consistent, yet fine-tuned,
flat backgrounds.}.  There are thus two possible scenarios.
\begin{itemize}
\item 
The superpotential $W$ grows as $e^{2\phi}$ or faster and the equation
of motion are not satisfied at the singularity. The only consistent
way out is to resolve the singularity, for instance by introducing a
second brane located at $y_s$. In that case fine-tuning of the CC is
restored as we introduce two more conditions analogous to
Eqs.~(\ref{bc1}) and (\ref{bc2}) or Eq.~(\ref{W0}) respectively, but do
not increase the number of free parameters~\cite{Forste:2000ps}.
\item
The superpotential grows as $e^{\nu\phi}$ with $\nu<2$, or slower. The
equations of motion are satisfied at the singularity, and there is no
need to resolve it. The price one pays is the adjustment of the
integration constant in Eq.~(\ref{W}). In this case we lose one of our
parameters needed to satisfy the boundary condition Eq.~(\ref{W0}),
resulting in a fine-tuning of the brane tension.
\end{itemize}
It is important to realize that either fine-tuning precisely corresponds to the fine-tuning of the CC.  
In the second possibility above this is particularly obvious: the superpotential is completely specified by the bulk potential and the boundary condition at $\phi\to\infty$.  
Eq.~(\ref{W0}) is then simply the minimization of the 4D potential
\be
V_4(\phi)\equiv\lambda(\phi)-6 W(\phi)
\label{V4}
\ee
under the condition that $V_4(\phi)$ vanishes at the minimum $\phi=\phi_0$. In fact the brane potential $\lambda(\phi)$ should be determined by physics localized at the UV brane interacting with the (dilaton) field $\phi$. For example if the Standard Model Higgs field $H$ is localized at the UV brane it will generate a brane potential as $\lambda(\phi,H)$ which will in turn provide the effective brane potential $\lambda(\phi)\equiv \lambda(\phi,\langle H\rangle)$ after electroweak symmetry breaking. So after the electroweak phase transition~\footnote{Other phase transitions, as e.g.~the QCD phase transition, should have a similar effect.} there will be a $\phi$-dependent vacuum energy which will require re-tuning the cosmological constant to zero and possibly a shift in the minimum of Eq.~(\ref{V4}).

What matters to us here is that there exist consistent solutions to the equations of motion in the full closed interval $[0,y_s]$ that, although requiring a fine-tuning of the CC,  do not demand the introduction of a second brane or any other means of resolving the singularity.

\section{A Model of Soft-Wall Stabilization}
\label{model}

Our goal in this section is to find solutions to the 5D scalar-gravity system that:
\begin{enumerate}
\item
require only one brane at $y=0$,
\item
behave as $AdS_5$ near that brane,
\item
have a soft wall at a value $y=y_s$, i.e.~give rise to finite volume,
\item
possess a mass gap or level spacing hierarchically smaller than the
Planck scale without fine-tuning of parameters.
\end{enumerate}
The last requirement might be called the problem of soft-wall
stabilization.
We will now show that all these requirements can be met by starting
from the simple superpotential
\be
W=k(1+e^{\nu\phi})\,,
\label{ourW}
\ee
where $k$ is some arbitrary dimensionful constant of the order of the 5D Planck scale, and $\nu<2$. The solution can be
immediately written down
\bea
A(y)&=&k y-\frac{1}{\nu^2}\log\left(1-\frac{y}{y_s}\right)\nn ,\\
\phi(y)&=&-\frac{1}{\nu}\log[\nu^2k(y_s-y)]\,.
\label{backgroundsolution}
\eea
At the point $y=y_s$ we encounter a naked curvature singularity as
explained in Section~\ref{general}. For $y\ll y_s$, i.e.~near the boundary at $y=0$, the
geometry is $AdS_5$.

The bulk potential which corresponds to the superpotential (\ref{ourW}) is given by
\be
V(\phi)=(3 k^2\nu^2-12 k^2)e^{2\nu\phi}-24 k^2e^{\nu\phi}-12 k^2
.
\label{potential}
\ee
\begin{itemize}
\item
For $\nu\leq 2$ the potential is bounded from above. More precisely it satisfies the condition
\be
V(\phi[y])\leq V(\phi_0)
,
\label{condition}
\ee
necessary~\cite{Gubser:2000nd} for the corresponding bulk
geometry to support finite temperature in the form of black hole
horizons~\footnote{A characteristic feature of physical
singularities~\cite{Gubser:2000nd}.}. Moreover for $\nu<2$, as we have
seen in the previous section, the equations of motion are satisfied at
the singularity and there is no need to resolve it.

\item
For $\nu>2$ the equations of motion are not satisfied at the
singularity and the latter would need to be resolved to fine-tune to
zero the four-dimensional cosmological constant. Finally the potential
is not bounded from above and finite temperature is not supported in the dual theory.
\end{itemize}

\begin{center}
\begin{figure}[thb]
\centering
\begin{psfrags}
\psfrag{absphi0}[tc][tc]{ $\vert \phi_0 \vert$}%
\psfrag{Boxkys30}[cc][cc]{ $k y_s = 30$}%
\psfrag{log10rhok}[bc][bc]{ $\log_{10} ( \rho/k)$}%
\psfrag{Nu1}[cc][cc]{ $\nu =1$}%
\psfrag{Nu2}[cc][cc]{ $\nu =2$}%
\psfrag{Nu}[tc][tc]{ $\nu $}%
\psfrag{S11}[tc][tc]{ $1$}%
\psfrag{S121}[tc][tc]{ $1.2$}%
\psfrag{S141}[tc][tc]{ $1.4$}%
\psfrag{S151}[tc][tc]{ $1.5$}%
\psfrag{S161}[tc][tc]{ $1.6$}%
\psfrag{S181}[tc][tc]{ $1.8$}%
\psfrag{S21}[tc][tc]{ $2$}%
\psfrag{S251}[tc][tc]{ $2.5$}%
\psfrag{S31}[tc][tc]{ $3$}%
\psfrag{S351}[tc][tc]{ $3.5$}%
\psfrag{S41}[tc][tc]{ $4$}%
\psfrag{W0}[cr][cr]{ $0$}%
\psfrag{Wm12}[cr][cr]{ $-10$}%
\psfrag{Wm1342}[cr][cr]{ $-13.4$}%
\psfrag{Wm1362}[cr][cr]{ $-13.6$}%
\psfrag{Wm1382}[cr][cr]{ $-13.8$}%
\psfrag{Wm1422}[cr][cr]{ $-14.2$}%
\psfrag{Wm142}[cr][cr]{ $-14$}%
\psfrag{Wm1442}[cr][cr]{ $-14.4$}%
\psfrag{Wm152}[cr][cr]{ $-15$}%
\psfrag{Wm22}[cr][cr]{ $-20$}%
\psfrag{Wm252}[cr][cr]{ $-25$}%
\psfrag{Wm32}[cr][cr]{ $-30$}%
\psfrag{Wm51}[cr][cr]{ $-5$}%
\psfrag{x0}[tc][tc]{ $0$}%
\psfrag{x11}[tc][tc]{ $1$}%
\psfrag{x2}[tc][tc]{ $0.2$}%
\psfrag{x4}[tc][tc]{ $0.4$}%
\psfrag{x6}[tc][tc]{ $0.6$}%
\psfrag{x8}[tc][tc]{ $0.8$}%
\psfrag{y0}[cr][cr]{ $0$}%
\psfrag{y11}[cr][cr]{ $1$}%
\psfrag{y2}[cr][cr]{ $0.2$}%
\psfrag{y4}[cr][cr]{ $0.4$}%
\psfrag{y6}[cr][cr]{ $0.6$}%
\psfrag{y8}[cr][cr]{ $0.8$}%
\includegraphics{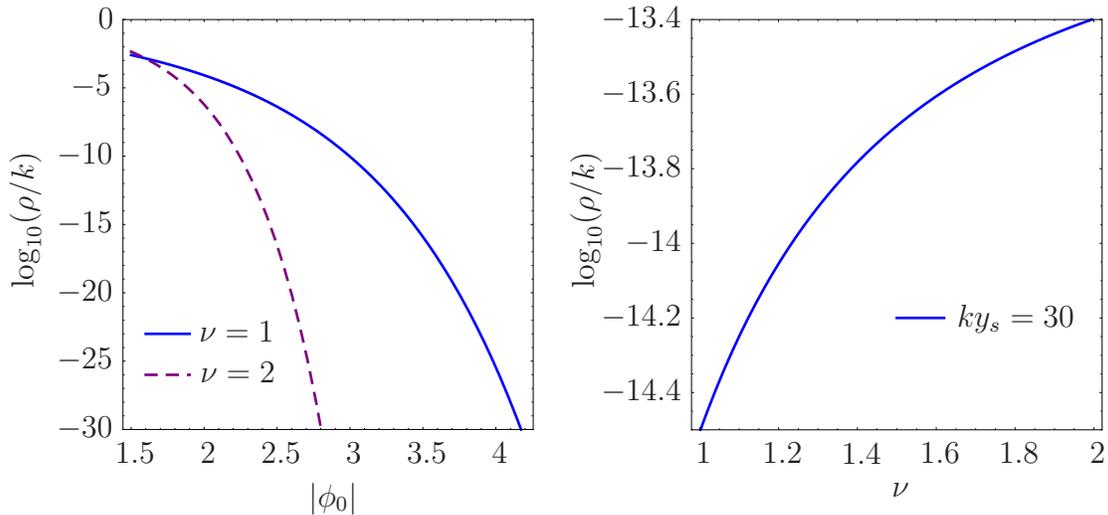}
\end{psfrags} 
\caption{\it Plot of $\log_{10}(\rho/k)$ as a function of $|\phi_0|$ for $\nu=1$ and $\nu=2$ [left panel], and as a function of $\nu$ for $k y_s = 30$  (this value will be used in following plots) [right panel].
}
\label{FT}
\end{figure}
\end{center}

The location of the singularity depends exponentially on the brane
value of $\phi$,
\be
k y_s=\frac{1}{\nu^2} e^{-\nu \phi_0}.
\label{kys}
\ee
As we will see in the next section the relevant mass scale for the 4D
spectrum is not the inverse volume but rather the ``warped down'' quantity
\be
\rho\equiv k(k y_s)^{-1 / \nu^2 }e^{-k y_s}\,.
\label{rho}
\ee
All we need in order to create the electroweak hierarchy is thus
$\phi_0<0$ but otherwise of order unity.  This can be achieved with a
fairly generic brane potential, for instance by chosing a suitable
$\lambda(\phi)$ such that the second of Eq.~(\ref{W0}) is satisfied
for our superpotential~\footnote{In order to satisfy the first of
Eq.~(\ref{W0}) we still need a fine-tuning, for instance by adding a
$\phi$ independent term to $\lambda(\phi)$. This is precisely the
tuning of the 4D CC discussed above that of course has nothing to do
with the electroweak hierarchy we want to explain here.}.  
For negative $\phi_0$, the ratio of scales $k/\rho$ exhibits a double exponential behaviour
\be
\log\, \frac{k}{\rho}\sim \frac{e^{\nu(-\phi_0)}}{\nu^2}   +\dots
,
\label{rhok}
\ee
 and we can create a huge hierarchy with very little fine-tuning. In Fig.~\ref{FT} we
plot $\rho/k$ as a function of $|\phi_0|$ for different values of $\nu$ and also as a function of $\nu$ for a fixed value $k y_s=30$ which generates a hierarchy of about fourteen orders of magnitude.

A comment about the choice of our superpotential is in order here. Its
particular form, Eq.~(\ref{ourW}), guarantees full analytic control over our
solution. A more detailed analysis of other possibilities will be postponed to Section~\ref{others}.

It will be useful in the following to define the metric also in conformally flat coordinates defined by the line element
\be
ds^2=e^{-2A(z)}(dx^\mu dx^\nu\eta_{\mu\nu}+dz^2) .
\label{conformal}
\ee
where $A(z)\equiv A[y(z)]$, the relationship between $z$ and $y$ coordinates being given by $\exp[A(y)]dy=dz$.
One easily finds, for $\nu>0$, that
\be
\rho(z-z_0)=
\Gamma(1-1/\nu^2,ky_s-ky)-\Gamma(1-1/\nu^2,ky_s)
,
\label{rel}
\ee
where $z_0$ corresponds to the location of the UV brane that we assume to be at $z_0=1/k$ and $\Gamma(a,x)$ is the incomplete gamma function. Since we are taking $e^{k y_s}\gg1$ and hence $k/\rho\gg 1$ we can approximate $\Gamma(1-1/\nu^2,k y_s)\simeq \rho/k$ and (\ref{rel}) simplifies to
\be
\rho z\simeq\Gamma(1-1/\nu^2,ky_s-ky)\,. \label{relsimp}
\ee
For $\nu>1$ the singularity at $y_s$ translates into a singularity at $z_s$ given by
\be
\rho z_s\simeq%
\Gamma(1-1/\nu^2)\,.
\label{relsing}
\ee
For $0<\nu\leq1$ the singularity at $y_s$ translates into a singularity at $z_s\to\infty$. 

As we will see in the next section the case $0\leq \nu<1$ provides continuous spectra without any mass gap~\footnote{The case $\nu=0$ is just the RS2 model~\cite{RS2} with a constant dilaton $\phi$.}, i.e.~typically it leads from a 4D perspective to unparticles~\cite{Georgi:2007ek}. The case $\nu=1$ corresponds to continuous spectra with a mass gap provided by $\rho$ in Eq.~(\ref{rhok}) leading in 4D to unparticles with mass gaps~\cite{Cacciapaglia:2008ns,Falkowski:2008yr}. Finally the case $1<\nu<2$ corresponds to discrete spectra with typical level spacing controlled by $\rho$, as in AdS models with two branes~\cite{RS1}.

\section{Fluctuations and the 4D Spectrum}
\label{fluctuations}
In this section we study the fluctuations of the metric and scalar around the classical background solutions. A general ansatz to describe all gravitational excitations of the model is, with the appropiate gauge choice~\cite{Csaki:2000zn}
\begin{gather}
 \phi(x,y) = \phi(y) + \varphi(x,y) , \\
 ds^2 = e^{-2A(y) - 2F(x,y)}(\eta_{\mu\nu} + h^{\mathrm{TT}}_{\mu\nu}) dx^\mu dx^\nu + (1 + G(x,y))^2 dy^2 ,
\end{gather}
where $\phi(y)$ is the background solution given in Eq.~\eqref{backgroundsolution}. The Einstein equations that arise from this ansatz have the spin-two fluctuations decoupled from the spin-zero fluctuations, so we can proceed to study them independently.
\subsection{The Graviton}

Let us first consider the graviton as the transverse traceless fluctuations of the metric
\begin{equation}
 ds^2 = e^{-2A(y)} (\eta_{\mu\nu} + h_{\mu\nu} (x,y) ) dx^2 + dy^2,
\end{equation}
where $h_{\mu}^{\ \mu} = \partial_\mu h^{\mu\nu} = 0$. In order to respect the orbifold symmetry and to keep the possibility of a constant profile zero mode, we will consider $h(y) = h(-y)$ which leads to the boundary condition at the brane $h'(0)=0$. The part of the action quadratic in the  graviton fluctuations becomes
\begin{multline}
 S = \int d^4 x\, d y\, \sqrt{-g} R\\  \rightarrow -\frac14 \int d^4 x\, d y\, e^{-2 A(y)} \left(  \partial_\rho h_{\mu\nu} \partial^\rho h^{\mu\nu} + e^{-2A(y)}  \partial_y h_{\mu\nu} \partial_y h^{\mu\nu}\right) .
 \label{actionh}
\end{multline}
Using the ansatz
\begin{equation}
h_{\mu\nu} (x,y) =  h^{}_{\mu\nu} (x) h^{}(y) ,
\end{equation}
one can obtain the equation of motion for the wavefunctions $h(y)$, which is given by
\begin{equation}
 h''(y) - 4 A'(y) h'(y) + e^{2A(y)} m^2 h(y) = 0  .
\label{GEOMy}
\end{equation}
After an integration by parts in \eqref{actionh}, one finds an additional equation due to boundary terms at $y=y_s$,  
\begin{equation}
 e^{-4 A(y_s)} h'(y_s)  = 0 .
\label{GBCy}
\end{equation}
In addition, one has to impose that the solutions are normalizable, i.e.
\begin{equation}
 \int_{0}^{y_s} dy \, e^{-2A(y)} h^2(y) < \infty .
\end{equation}

It is now convenient to change to conformally flat coordinates, as defined in~\eqref{rel}. In this frame, rescaling the field by $ \tilde h(z) = e^{-3 A(z)/2} h(z)$, Eq.~\eqref{GEOMy} can be written as a Schroedinger-like equation, 
\begin{equation}
 - \ddot {\tilde{ h}}(z) +  V_{h}(z) \tilde h(z) = m^2 \tilde h(z) ,
\label{GSchroedinger}
\end{equation}
where a dot denotes derivation with respect to $z$, and the potential is given by
\begin{equation}
  V_{h} (z) = \frac{9}{4} \dot A^2 (z) - \frac{3}{2} \ddot A(z) 
\end{equation}
The boundary equations are written in the $z$-frame as
\begin{equation}
 \left.e^{-3 A(z)} \dot h(z) \right\vert_{z_0,z_s} 
 = 
 \left.e^{-3 A(z)/2} \left( \dot{\tilde h} (z) + \frac32 \dot A(z) \tilde h(z) \right) \right\vert_{z_0,z_s} 
 = 0 ,
 \label{Gconditionszs}
\end{equation}
and the normalizability condition is
\begin{equation}
 \int_{z_0}^{z_s} dz \, e^{-3A(z)} h^2(z) = \int_{z_0}^{z_s} dz \, \tilde{h}^2(z)  < \infty
 .
 \label{Gnormalizzs}
\end{equation}

\begin{figure}[ht]
\begin{psfrags}
\psfrag{S1k}[tc][tc]{$1/k$}%
\psfrag{Szs}[tc][tc]{$z_s$}%
\psfrag{W0}[cr][cr]{ $0$}%
\psfrag{Wmg}[cr][cr]{$m_g$}%
\psfrag{WV0}[cr][cr]{$V_0$}%
\psfrag{z}[tc][tc]{$z$}%

\centering
\subfigure[$0<\nu<1$]{
\includegraphics{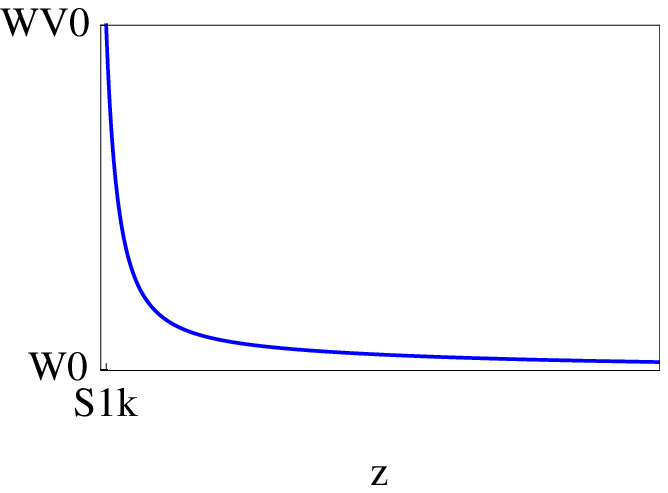}
\label{fig:potential<1}
}
\subfigure[$\nu=1$]{
\includegraphics[scale=1]{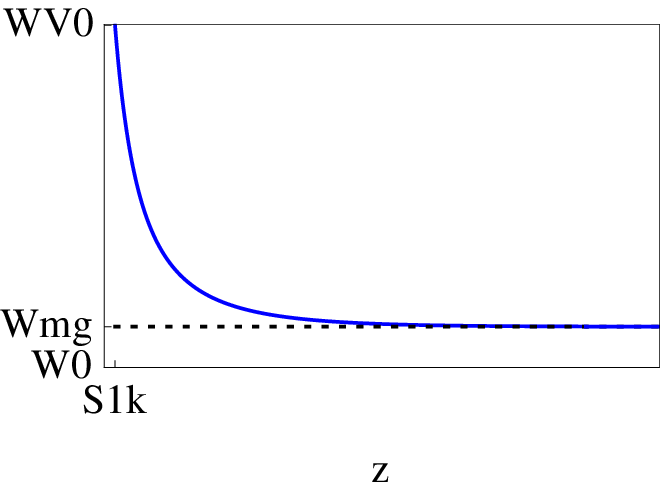}
\label{fig:potential=1}
}
\subfigure[$1<\nu<\sqrt{5/2}$]{
\includegraphics{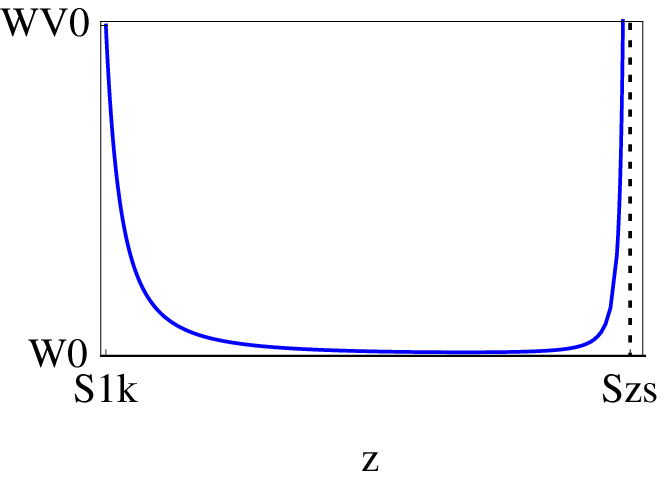}
\label{fig:potential>1}
}
\subfigure[$\nu>\sqrt{5/2}$]{
\includegraphics{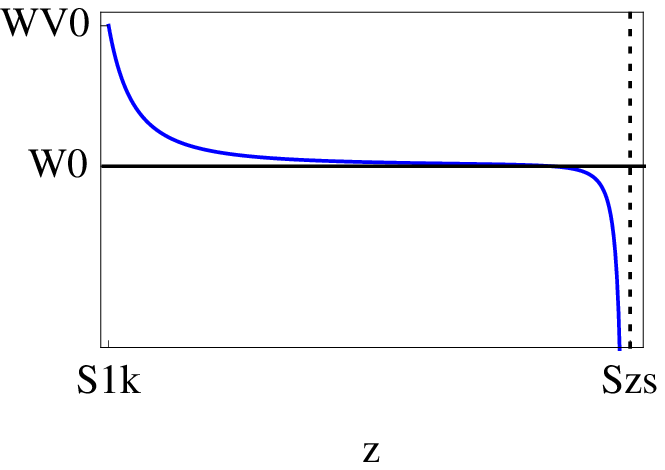}
\label{fig:potential>5/2}
}
\caption{\it Behaviour of $V_{h}(z)$ for different values of $\nu$. Here, $V_0\equiv V_h(1/k)$. For the radion, $V_{F}(z)$ has the same behaviour with the exception that Fig.~\ref{fig:potential>1} applies for all $\nu>1$.  }
\label{fig:potential}
\end{psfrags}
\end{figure}

In the case of study, it is only possible to obtain an analytic expression for the potential in the $y$-frame, where it reads
\begin{equation}
  V_{h} (z[y]) = \frac{3 e^{-2 k y} \left(1-\frac{y}{y_s}\right)^{\frac{2}{\nu^2}} \left[\,5 \nu^4 k^2  (y-y_s)^2
 -10 \nu ^2 k
   (y-y_s) - 2 \nu^2 +5\,\right]}{4 \nu ^4 (y-y_s)^2}\, .
 \label{GVhat}
\end{equation}
It is however possible to invert numerically the coordinate change \eqref{rel}, and so to plot \eqref{GVhat}. Its behaviour for different values of $\nu$ is shown in Fig.~\ref{fig:potential}. One can distinguish three possible situations~\footnote{Similar potentials were considered in Ref.~\cite{Freedman:1999gk}.}:
\begin{itemize}
 \item $\nu<1$ [Fig.~\ref{fig:potential<1}] In this case $z$ extends to infinity where $ V_{h} \rightarrow 0$. The mass spectrum is continuous from $m=0$, leading to unparticles without a mass gap. However, conformal symmetry is broken due to the occurrence of the scale $y_s$.
 \item $\nu=1$ [Fig.~\ref{fig:potential=1}] $z$ also extends to infinity but $ V_{h} \rightarrow (9/4)\rho^2$. This leads to unparticles with a mass gap $m_{\mathrm{g}} = (3/2) \rho$. 
 \item $\nu>1$ [Figs.~\ref{fig:potential>1} and \ref{fig:potential>5/2}] $z_s$ is finite and thus the mass spectrum is discrete. The potential diverges at $z_s$ changing sign at $\nu^2 = 5/2$, but this does not have observable consequences in the mass spectrum as we will see. 
\end{itemize}

Equations \eqref{GEOMy} and \eqref{GSchroedinger} do not have analytic solutions.  However, for $\nu>1$ one can find approximations for the wavefunction in the regions near the brane and near the singularity.  Let us first consider the region near the brane ($ky \simeq 0$). Assuming $k y_s \gg 1$ the potential \eqref{GVhat} is approximated as
\begin{equation}
  V_{h} \vert_{y\simeq 0} \simeq \frac{15\, k^2}{4} e^{-2 k y} \simeq \frac{15}{4} \frac{1}{z^2} , 
 \label{GVhatapproxy0}
\end{equation}
where the coordinate change is given by \eqref{relsimp}, which is approximated for $\nu > 1$ as
\begin{equation}
 k z  \simeq  e^{ky} .
 \label{coordapprox0}
\end{equation}
One can see that \eqref{GVhatapproxy0} corresponds to an AdS metric. With this approximated potential, Eq.~\eqref{GSchroedinger} is solved by
\begin{equation}
 \tilde h(z) \vert_{z\simeq z_0} = c_{1} \sqrt{k z} J_2 (m z ) + c_{2} \sqrt{k z} Y_{2} (m z) .
\end{equation}
The two coefficients can be determined by the normalization and the boundary condition \eqref{Gconditionszs} at $z_0$, which yields
\begin{equation}
 \frac{c_2}{c_1} = -\frac{J_1(m/k)}{Y_1(m/k)} \sim \left( \frac{m}{k} \right)^2 \simeq 0 ,
\end{equation}
since we expect the first mass modes to be of order $m \simeq (z_s - z_0)^{-1}$, and in our approximation $k (z_s - z_0) \gg 1$. 

Let us now move on to consider the region next to the singularity ($y\simeq y_s$). In this case the potential is approximated by
\begin{equation}
  V_{h} \vert_{y\simeq y_s} \simeq \frac{3(5 - 2\nu^2)}{4 \nu^4} 
  \frac{ \rho^2 }{ [k(y_s-y)]^{2-2/\nu^2}} 
\simeq
\frac{3(5 - 2\nu^2)}{4(1 - \nu^2)^2} \frac{ 1 }{ (z_s-z)^{2}}
,
\label{GVhatapprox1}
\end{equation}
where we have used that, for $\nu>1$, the coordinate change \eqref{rel} is approximated by
\begin{equation}
 \rho(z_s - z) = \frac{\nu^2}{\nu^2 -1} [k (y_s - y)]^{1-1/\nu^2} .
 \label{coordsingularity}
\end{equation}
With this approximation, Eq.~\eqref{GSchroedinger} yields the solution
\begin{equation}
 \tilde h (z) = c_J  \sqrt{m \Delta z} J_\alpha ( m \Delta z) + c_Y  \sqrt{ m \Delta z} Y_\alpha ( m \Delta z ),
 \label{GsolutionII}
\end{equation}
where
\begin{equation}
 \alpha = \frac{4-\nu^2}{2(\nu^2-1)}
\end{equation}
and $\Delta z \equiv z_s - z $. The two integration constants can be obtained by imposing the boundary condition at the singularity and normalizability and by matching this solution to the solution for the intermediate region between the  brane and the singularity. Near the singularity \eqref{GsolutionII} behaves like
\begin{equation}
 \tilde h (z) 
 \sim
  c^{(1)}_J (\Delta z)^{3/(2\nu^2-2)} + c^{(2)}_J (\Delta z)^{(4\nu^2 -1)/(2\nu^2 -2)} + 
 c^{(1)}_Y (\Delta z)^{(2\nu^2 - 5)/(2\nu^2 - 2)} 
,
\label{Gsolutionzs}
\end{equation}
where numerical factors are being absorbed in the constants $c_i$. We have included the next to leading order in the expansion of $J_\alpha$ as we need it for computing the boundary condition, which reads
\begin{align}
 e^{-3A/2} \left( \dot{\tilde h} (z) + \frac{3}{2} \dot A \tilde h \right) 
 \sim 
 c^{\prime(2)}_J (\Delta z)^{(\nu^2 + 2)/(\nu^2 -1) }
+ 
 c^{\prime(1)}_Y (\Delta z)^0 .
 \label{Gbczs}
\end{align}
Again numerical factors have been absorbed in $c'_i$. Note that the boundary condition is only satisfied when $c_Y = 0$, and that this condition also ensures that the solution \eqref{Gsolutionzs} is normalizable when $\nu^2 < 2$.

The boundary conditions provide the quantization of the mass eigenstates for $\nu>1$. In order to compute the mass spectrum for the graviton one should match the solutions at the ends of the space with a solution for the intermediate region. Unfortunately, for the parameter range we are interested in we do not have good analytic control for this region. However we can extract a generic property of the spectrum by looking at the potential Eq.~(\ref{GVhat}) and using the form of the coordinate transformation Eq~(\ref{relsimp}) to deduce that, assuming $e^{k y_s} \gg 1$, the potential has the form~\footnote{This behaviour can actually be seen in the limiting cases Eq.~(\ref{GVhatapproxy0}) and Eq.~(\ref{GVhatapprox1}).}
\be
V_h(z)=\rho^2\, v_h(\rho z)\,,
\label{scaling1}
\ee
where $v_h$ is some dimensionless function of the dimensionless variable $\rho z$. 
In other words we have eliminated the two scales $k, y_s$ in favour of the single scale $\rho$ given in Eq.~(\ref{rho}). The spectrum is therefore of the form
\be
m_n(\nu,k,y_s)=\mu_n(\nu)\,\rho(\nu,k,y_s)\,,
\label{scaling2}
\ee
where the pure numbers $\mu_n$ only depend on the parameter $\nu$ but not on the parameters $k$ or $y_s$. 

\begin{figure}[t]
\centering
\begin{psfrags}
\psfrag{mRhoNu}[bc][bc]{ $m/\rho(\nu)$}%
\psfrag{Nu}[tc][tc]{ $\nu $}%
\psfrag{S11}[tc][tc]{ $1$}%
\psfrag{S121}[tc][tc]{ $1.2$}%
\psfrag{S141}[tc][tc]{ $1.4$}%
\psfrag{S161}[tc][tc]{ $1.6$}%
\psfrag{S181}[tc][tc]{ $1.8$}%
\psfrag{S21}[tc][tc]{ $2$}%
\psfrag{W0}[cr][cr]{ $0$}%
\psfrag{W122}[cr][cr]{ $12$}%
\psfrag{W12}[cr][cr]{ $10$}%
\psfrag{W142}[cr][cr]{ $14$}%
\psfrag{W21}[cr][cr]{ $2$}%
\psfrag{W41}[cr][cr]{ $4$}%
\psfrag{W61}[cr][cr]{ $6$}%
\psfrag{W81}[cr][cr]{ $8$}%
\includegraphics{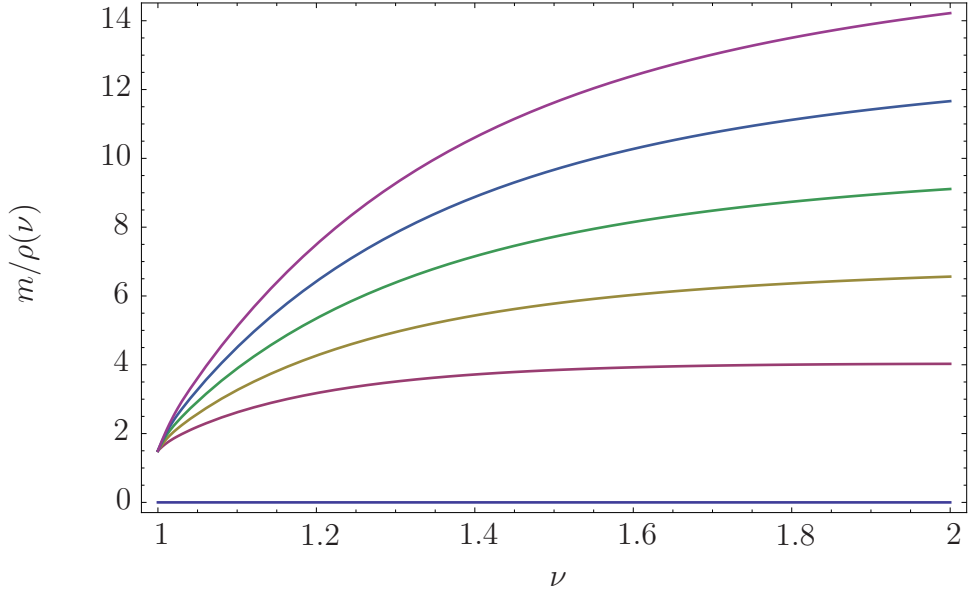}
\end{psfrags} 
\caption{\it Mass modes for the graviton, computed for $k y_s > 4$~\textsuperscript{\ref{footnotec}}. The massless $(n=0)$ and the first 5 massive modes $(n=1,\ldots,5)$ are shown.}
\label{fig:massesG}
\end{figure}

\begin{figure}[t]
\centering
\begin{psfrags}
\psfrag{n0}[cc][cc]{ $\text{n=0}$}%
\psfrag{n1}[cc][cc]{ $\text{n=1}$}%
\psfrag{n2}[cc][cc]{ $\text{n=2}$}%
\psfrag{S0}[tc][tc]{ $0$}%
\psfrag{S11}[tc][tc]{ $1$}%
\psfrag{S2}[tc][tc]{ $0.2$}%
\psfrag{S4}[tc][tc]{ $0.4$}%
\psfrag{S6}[tc][tc]{ $0.6$}%
\psfrag{S8}[tc][tc]{ $0.8$}%
\psfrag{SqrtBoxBoxA}[Bc][Bc]{ $\sqrt{z_s} \, \tilde{h}(z)$}%
\psfrag{SqrtBoxBox}[cr][cr]{ $\sqrt{z_s} \, \tilde{h}(z)$}%
\psfrag{W0}[cr][cr]{ $0$}%
\psfrag{W11}[cr][cr]{ $1$}%
\psfrag{W151}[cr][cr]{ $1.5$}%
\psfrag{W5}[cr][cr]{ $0.5$}%
\psfrag{Wm11}[cr][cr]{ $-1$}%
\psfrag{Wm151}[cr][cr]{ $-1.5$}%
\psfrag{Wm5}[cr][cr]{ $-0.5$}%
\psfrag{zzsA}[Bc][Bc]{ $z/z_s$}%
\psfrag{zzs}[tc][tc]{ $z/z_s$}%
\includegraphics{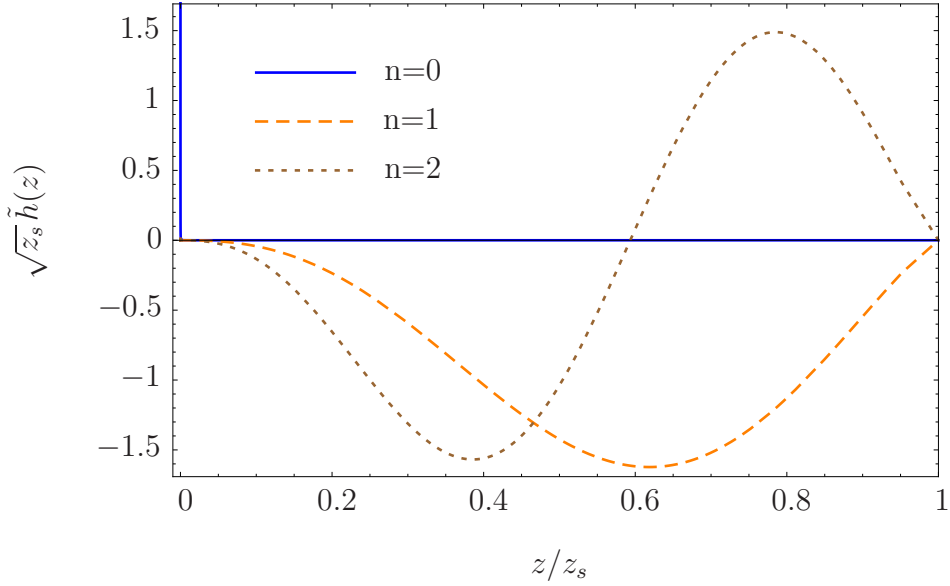}
\end{psfrags} 
\caption{\it KK graviton profiles in the $z$ frame for $k y_s = 30$ and $\nu=3/2$, using the normalization $\int dz \, \tilde h^2 =1$. The massless mode ($n=0$) is peaked near the brane. The two first massive modes ($n=1,2$) are also shown. The zero mode becomes more peaked near the brane in comparison to the massive modes as $ky_s$ increases.}
\label{fig:gprofile}
\end{figure}

Moreover one can find an expression for the spacing of the mass eigenstates by approximating the potential as an infinite well, which is valid for $m^2 \gg  V_{h}$. The result of this approximation is
\begin{equation}
 \Delta m \simeq   \frac{\rho\, \pi}{\Gamma\left( 1 - 1/\nu^ 2 \right)}=
\frac{\pi}{z_s} \,.
 \label{GDeltam}
\end{equation}
Note that the mass spectrum is linear ($m_n \sim n$), and that as one approaches $\nu=1$
\begin{equation}
  \lim_{\nu \to 1} \ \Delta m=0 ,
\end{equation}
recovering the expected continuous spectrum at this value (for $\nu<1$ the spectrum is continuous too, since \eqref{GDeltam} is only valid for $\nu>1$). The numerical result for the mass eigenvalues is shown in Fig.~\ref{fig:massesG} where these behaviours can be observed. Some profiles for the graviton computed numerically using the equation of motion \eqref{GEOMy} and the boundary conditions \eqref{GBCy} are shown in Fig.~\ref{fig:gprofile}.

\footnotetext{Numerically one finds that the scaling property in Eq.~\eqref{scaling2} ceases to be valid for $k y_s \lesssim 3$, as discrepancies from this behavior become greater than $1\%$.  \label{footnotec}}

\subsection{The Radion-Scalar system}
Now we consider the spin-zero fluctuations of the system. This is
\begin{gather}
 \phi(x,y) = \phi(y) + \varphi(x,y) , \\
 ds^2 = e^{-2A(y) - 2F(x,y)}\eta_{\mu\nu} dx^\mu dx^\nu + (1 + G(x,y))^2 dy^2 .
\end{gather}
With an appropiate gauge choice, the equations of motion for the $y$-dependent part of the KK modes form a coupled system with only one degree of freedom. The derivation of the equations is given with detail in \cite{Csaki:2000zn}, and the result is
\begin{gather}
 F'' - 2 A' F' - 4 A'' F - 2 \frac{\phi''}{\phi'} F' + 4 A' \frac{\phi''}{\phi'} F = - m^2 e^{2A} F ,\\
 \phi' \varphi =  F' - 2 A' F , \\
 G = 2F .
 \label{Feqmotion}
\end{gather}
The boundary equations on the brane depend on the brane tension $\lambda(\phi)$. The precise form of the dependence can be found in \cite{Csaki:2000zn}. At the singularity, similarly to the graviton case, one gets the boundary equation
\begin{equation}
 e^{-4A(y)} \varphi'(y)\vert_{y_s} = 0 ,
\label{Fboundary}
\end{equation}
and the normalizability condition
\begin{equation}
 \int_{0}^{y_s} dy e^{-2A(y)} \varphi^2 (y) = \int_{z_0}^{z_s} dz \tilde{\varphi}^2(z)  < \infty ,
\label{Fnorm}
\end{equation}
where the field has been rescaled by $\tilde \varphi (z) \equiv e^{-3A/2} \varphi (z)$.

It is convenient, as for the graviton,  to use conformally flat coordinates. Rescaling the field by $\tilde F(z) = e^{-3A(z)/2}  F(z)/{\dot \phi(z)}$, \eqref{Feqmotion} can be written as the Schroedinger equation
\begin{equation}
 -\ddot{\tilde F}(z) +  V_{F} (z) \tilde F(z) = m^2 \tilde F(z) ,
 \label{Fschroedinger}
\end{equation}
where
\begin{equation}
 V_{F} (z) = \frac94 \dot A ^2 + \frac52 \ddot A - \dot A \frac{\ddot \phi}{\dot \phi} - \frac{\dddot \phi}{\dot \phi}+ 2 \left( \frac{\ddot \phi}{\dot \phi}\right)^2 \ .
 \label{FVhatz}
\end{equation}
The relation between the rescaled field $\tilde F$ and the scalar field $\varphi$ is 
\begin{equation}
  \varphi (z) = e^{3A/2} \left[ \dot{ \tilde F} + \left( \frac{\ddot\phi}{\dot \phi} - \frac12 \dot A \right) \tilde F \right] .
 \label{Ftildevarphi}
\end{equation}

In the $y$-frame, Eq.~\eqref{FVhatz} is given by
\begin{equation}
  V_{F} (z[y]) = \frac{ e^{-2 k y} \left(1- \frac{y}{ y_s}\right)^{\frac{2}{\nu^2}} 
 \left[
 3 \nu^4 k^2  (y-y_s)^2
 + (-6\nu^2 + 8 \nu^4) k   (y-y_s)
 + 6\nu^2 + 3
 \right]}{4 \nu ^4 (y-y_s)^2}\,.
 \label{FVhat}
\end{equation}
This potential has similar form to the graviton potential, and the three situations presented above also apply for the radion (with the same mass gap for $\nu=1$). A difference is that this potential does not change the sign of divergence for $\nu>1$ but, as said before, this does not have any observable consequences.

\begin{figure}[t]
\centering
\begin{psfrags}
\psfrag{mRhoNu}[bc][bc]{   $m/\rho(\nu)$}%
\psfrag{Nu}[tc][tc]{   $\nu $}%
\psfrag{S11}[tc][tc]{   $1$}%
\psfrag{S121}[tc][tc]{   $1.2$}%
\psfrag{S141}[tc][tc]{   $1.4$}%
\psfrag{S161}[tc][tc]{   $1.6$}%
\psfrag{S181}[tc][tc]{   $1.8$}%
\psfrag{S21}[tc][tc]{   $2$}%
\psfrag{W0}[cr][cr]{   $0$}%
\psfrag{W1252}[cr][cr]{   $12.5$}%
\psfrag{W12}[cr][cr]{   $10$}%
\psfrag{W152}[cr][cr]{   $15$}%
\psfrag{W251}[cr][cr]{   $2.5$}%
\psfrag{W51}[cr][cr]{   $5$}%
\psfrag{W751}[cr][cr]{   $7.5$}%
\includegraphics{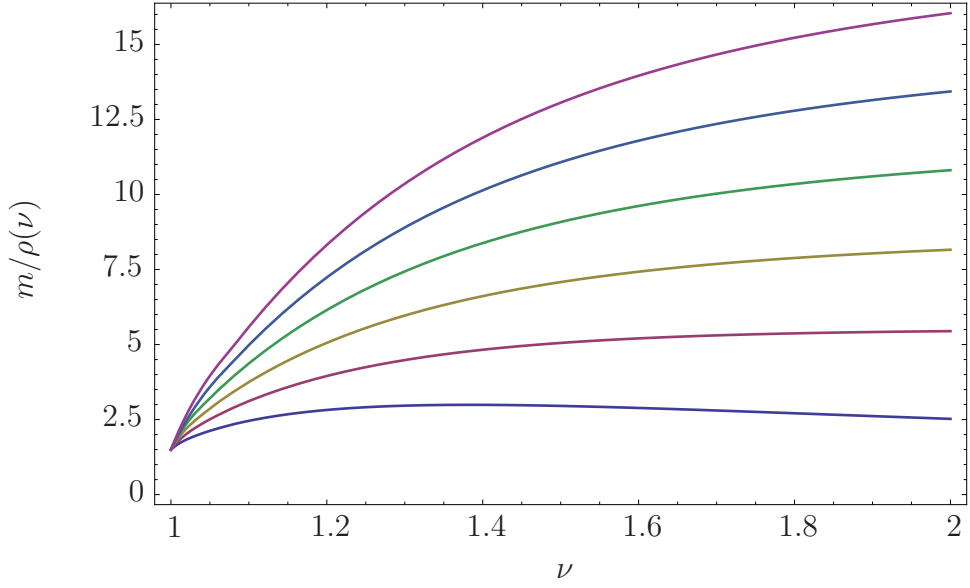}
\end{psfrags} 
\caption{\it Mass modes for the radion, computed for values of $k y_s > 4$ \textsuperscript{\ref{footnotec}}. The first 6 massive modes ($n=0,\ldots,5$) are shown.}
\label{fig:massesR}
\end{figure}

\begin{figure}[t]
\centering
\begin{psfrags}
\psfrag{n0}[cc][cc]{   $\text{n=0}$}%
\psfrag{n1}[cc][cc]{   $\text{n=1}$}%
\psfrag{n2}[cc][cc]{   $\text{n=2}$}%
\psfrag{S0}[tc][tc]{   $0$}%
\psfrag{S11}[tc][tc]{   $1$}%
\psfrag{S2}[tc][tc]{   $0.2$}%
\psfrag{S4}[tc][tc]{   $0.4$}%
\psfrag{S6}[tc][tc]{   $0.6$}%
\psfrag{S8}[tc][tc]{   $0.8$}%
\psfrag{SqrtBoxBoxA}[Bc][Bc]{   $\sqrt{z_s} \, \tilde{\varphi}(z)$}%
\psfrag{SqrtBoxBox}[cr][cr]{   $\sqrt{z_s} \, \tilde{\varphi}(z)$}%
\psfrag{W0}[cr][cr]{   $0$}%
\psfrag{W11}[cr][cr]{   $1$}%
\psfrag{W21}[cr][cr]{   $2$}%
\psfrag{Wm11}[cr][cr]{   $-1$}%
\psfrag{Wm21}[cr][cr]{   $-2$}%
\psfrag{zzsA}[Bc][Bc]{   $z/z_s$}%
\psfrag{zzs}[tc][tc]{   $z/z_s$}%
\includegraphics{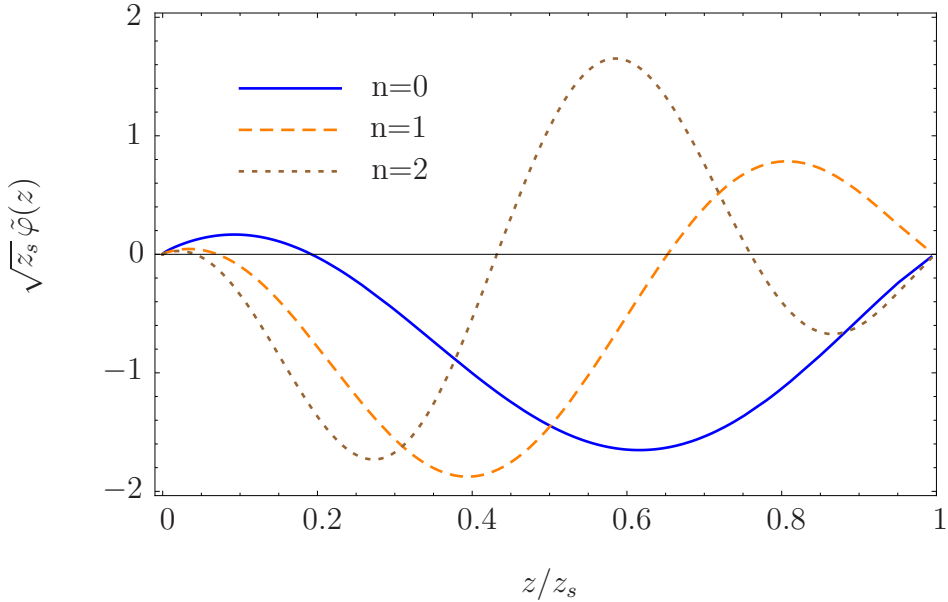}
\end{psfrags} 
\caption{\it KK profiles of the rescaled scalar fluctuations $\tilde\varphi(z)$ for
$k y_s = 30$ and $\nu=3/2$, using the normalization $\int dz \, \tilde \varphi^2 =1$.  The three first massive modes ($n=0,1,2$) are shown.}
\label{fig:rprofile}
\end{figure}

Let us now proceed to find the approximation for the wavefunction near the UV brane. Taking $k y \simeq 0$ and $k y_s \gg 1 $ and using \eqref{coordapprox0}, \eqref{FVhat} is given by
\begin{equation}
  V_{F} \vert_{y \simeq 0} \simeq \frac{3\,k^2}{4} e^{-2 k y} \simeq \frac34 \frac1{z^2} , 
\end{equation}
and hence the solution to \eqref{Fschroedinger} is
\begin{equation}
 \tilde F (z) \vert_{z\simeq z_0} = c_1 \sqrt{k z} J_1(mz) + c_2 \sqrt{k z} Y_1 (m z) . 
\end{equation}
The coefficients $c_i$ are to be determined using the boundary conditions at the brane \cite{Csaki:2000zn}. As an example, using the condition~\footnote{This condition holds for brane potentials satisfying $\partial^2 \lambda / \partial \phi^2 \gg 1$ \cite{Csaki:2000zn}.} $\varphi(y=0) = 0$  (which we will use for the numerical computation) yields
\begin{equation}
 \frac{c_2}{c_1} = - \frac{J_2(m/k)}{Y_2(m/k)} \simeq 0
.
\end{equation}

Next to the singularity, using \eqref{coordsingularity} the potential is approximated by
\begin{equation}
  V_{F} \vert_{y\simeq y_s} \simeq \frac{6\nu^2 + 3}{4 \nu^4} \frac{\rho^2}{ [k(y_s-y)]^{2-2/\nu^2}} 
\sim
\frac{6 \nu^2 + 3}{4(1 - \nu^2)^2} \frac{ 1 }{ (z_s-z)^{2}}
,
\end{equation}
that gives the solution
\begin{equation}
 \tilde F (z) = c_J  \sqrt{m \Delta z} J_\alpha ( m \Delta z) + c_Y  \sqrt{ m \Delta z} Y_\alpha ( m \Delta z ),
 \label{FsolutionII}
\end{equation}
with
\begin{equation}
 \alpha = \frac{2+\nu^2}{2\nu^2 -2} .
\end{equation}
The behaviour of this solution near the singularity is 
\begin{equation}
 \tilde F (z)  
\sim
  c^{(1)}_J  (\Delta z)^{(2\nu^2 +1)/(2\nu^2 -2)} + c^{(2)}_J (\Delta z)^{(6\nu^2 -3)/(2\nu^2 - 2)}   + 
 c^{(1)}_Y (\Delta z)^{-3/(2\nu^2 -2)}
.
\label{Rsolutionzs}
\end{equation}
Using \eqref{Ftildevarphi} we can compute the behaviour of the field and apply the normalizability condition \eqref{Fnorm}, 
\begin{equation}
\tilde{\varphi}(z) \sim  
 c^{\prime (1)}_J (\Delta z)^{3/(2\nu^2 -2)}  + 
 c^{\prime (1)}_Y (\Delta z)^{-(2\nu^2 + 1)/(2\nu^2 -2)}   ,
\end{equation}
and the boundary condition \eqref{Fboundary},
\begin{equation}
 e^{-3A(z)} \dot \varphi(z) \sim
 c^{\prime (2)}_J (\Delta z)^{(\nu^2+2)/(\nu^2 -1)}  + 
 c^{\prime\prime (1)}_Y (\Delta z)^{(-2\nu^2 + 2)/(\nu^2 - 1)}    .
\end{equation}
Again, the condition $c_Y = 0$ is sufficient to ensure both the fulfillment of the boundary conditions and the normalizability. The scaling of the mass eigenvalues Eq~(\ref{scaling2}) and the 
approximation \eqref{GDeltam} for the spacing of the mass modes also holds for the radion. The numerically obtained values for the masses are shown in Fig.~\ref{fig:massesR}. In comparison to the graviton mass modes of Fig.~\ref{fig:massesG}, note that the first mode for the radion is lighter than the first massive mode of the graviton. This can be understood recalling that the radion does not have a zero mode~{\footnote{One can in fact show that for $\nu\rightarrow \infty$, which we can only take if we resolve the singularity, the lightest mode tends to be massless, corresponding to the radion profile in \cite{Charmousis:1999rg}.}. 
Some profiles of the scalar fluctuations of the field $\tilde\varphi$ are shown in Fig.~\ref{fig:rprofile}.

\section{Other soft walls with a hierarchy}
\label{others}

The particular form of $W$, Eq.~(\ref{ourW}), guarantees full analytic control over our solution but may  seem a little ad hoc. It is natural to ask what are the essential ingredients of our stabilization mechanism and whether it is possible to generalize it to other potentials or superpotentials. The location of the singularity, and hence the size of the extra dimension, is given by
\be
y_s=\int_{\phi_0}^\infty\frac{d\phi}{W'(\phi)}\,.
\ee
Here and in the following we will assume that $W$ is a monotonically increasing function of $\phi$, i.e.~$W'(\phi)>0$.
The integral is finite whenever $W$ diverges faster than $W\sim \phi^2$.

However, the inverse volume $y_s^{-1}$ is, in general, not the 4D KK scale nor the mass gap as there might be a strong AdS warping near the UV brane. The KK scale 
is given by the inverse conformal volume $z_s^{-1}$ (when it is finite), calculated as
%
\be
z_s=\int_0^{y_s} e^{A(y)} dy.
\ee
It is easy to warp the geometry near the brane without affecting $y_s$ by adding a positive constant of $\mathcal O(k)$ to the superpotential, leading to
\be
A(y)\to A(y)+ky\,,\qquad{\rm for}\qquad W\to W+k\,
.
\label{trick}
\ee
Notice that $A(y)$ is a monotonically increasing function of $y$, such that 
\be
k z_s>e^{k y_s}\,.
\label{z_s}
\ee
One sees that the KK scale is warped down with respect to the compactification scale, a phenomenon well known in RS models with two branes~\cite{RS1}. In order to obtain, e.g., the TeV from the Planck scale we need
\be
k y_s=\int_{\phi_0}^\infty \frac{k}{W'(\phi)}\,d\phi\simeq 37\,.
\label{kys2}
\ee
This is not hard to achieve in a natural manner. In our model, Eq.~(\ref{ourW}), it works so well because the exponential behavior that was introduced for large values of $\phi$ is also valid at $\mathcal O(1)$ negative values and dominates the integral, leading to Eq.~(\ref{kys}). 

Moreover there are many cases where $z_s$ is infinite, even though $y_s$ is finite. There can still be mass gaps or even a discrete spectrum, but $z_s$ is clearly inadequate to characterize the energy levels. One such example is the case $W=ke^\phi$ that leads to a mass gap. Let us be slightly more general and consider the class of superpotentials given by
\be
W(\phi)= k\,e^{\phi}(\phi-\phi_1)^\beta\,,
\ee
with $\phi_1<\phi_0$.
This superpotential is monotonically increasing for $\beta\geq 0$ and has infinite $z_s$ for $\beta\leq\frac{1}{2}$, so we will assume $0\leq\beta\leq\frac{1}{2}$. The volume $y_s$ is approximately 
\be
ky_s \simeq e^{-\phi_0}\,,
\label{ys2}
\ee
so, again, $k y_s$ is (mildly) exponentially enhanced when $|\phi_0|=\mathcal O(1-10)$, $\phi_0<0$. In order to estimate the spectrum, we need the asymptotic behavior of the warp factor in conformally flat coordinates. For large $z$, it is given by
\be
A(z)\simeq (\rho z)^\frac{1}{1-2\beta}\,
,
\label{metricz}
\ee
where $\rho=\mathcal O(y_s^{-1})$. The coordinate change is given by
\be
z(y)=\int e^{A(y)} dy\,.
\ee
Using our trick of adding warping while keeping $y_s$ unchanged, Eq.~(\ref{trick}), we see that near the singularity 
\be
z(y)\to z_w(y)\simeq z(y)e^{k y_s}\,.
\ee
On the other hand, adding the warping leaves $A(y)$ nearly unchanged near the singularity (adding a constant $k y_s$ to infinity makes no difference).
Demanding thus $A_w(y)\simeq A(y)$ near $y=y_s$ leads to
\be
[\rho z(y)]^{1/(1-2\beta)}=[\rho_w z_w(y)]^{1/(1-2\beta)}
=[\rho_w z(y)e^{k y_s}]^{1/(1-2\beta)}
,
\ee
and hence
\be
\rho_w=\rho e^{-k y_s}\simeq\frac{e^{-k y_s}}{y_s}\,.
\ee
Combining this with Eq.~(\ref{ys2}) we find a strong suppression of  $\rho_w/k$ resulting just from $\mathcal O(1)$ numbers. The quantity $\rho_w$ sets the scale for the KK spectrum in this case. In fact metrics of the form Eq.~(\ref{metricz}) have been studied in Ref.~\cite{Batell2}. A WKB approximation shows that the spectrum can be approximated by
\be
m_n\simeq\rho_w\, n^{2\beta}\,.
\label{mn}
\ee
We see that $\rho_w$ indeed sets the scale of the 4D masses, which are hence parametrically suppressed with respect to $k$. The complete superpotential that accomplishes a hierarchy and leads to the spectrum Eq~(\ref{mn}) is
\be
W(\phi)=k(1+e^{\phi}[\phi-\phi_1]^\beta)\,.
\label{super}
\ee
In particular the case $\beta=1/4$ generates the linear Regge trajectory spectrum $m_n^2\simeq \rho_w^2 n$ appropriate for AdS/QCD models as in Ref.~\cite{AdS/QCD}.
In this case one would obtain the linear confinement behaviour of e.g.~$\rho$-mesons by considering an additional piece in our action $\int d^5 x \sqrt{-g}\, e^{-\frac{1}{2}\phi} \mathcal{L}_\mathrm{mesons}$. 
The fact that asymptotically $A(z)\sim\phi(z)\sim z^2$ guarantees that the resonances of the vector mesons follow the same linear law as the ones for the scalars and tensors.


Let us conclude this section by noting that there are certainly other ways to obtain the mild hierarchy $k y_s$, including moderate fine-tunings of parameters.
What is completely generic, though, is the fact that adding warping as in Eq.~(\ref{trick}) leaves $k y_s$ manifestly unchanged but suppresses the masses by an additional warp factor $e^{k y_s}$.

\section{Conclusions and Outlook}
\label{conclusions}
In this paper we have studied the stabilization of soft walls, i.e.~4+1 dimensional geometries with 4D Poincar\'e invariance, that are only bounded by a single three-brane but that nevertheless exhibit a finite volume for the extra dimensional coordinate. The second brane is typically replaced by a naked curvature singularity at a finite proper distance.
In particular we have studied how these soft walls arise in models with a single scalar field, and classified the type of models that can be realized as full solutions to the Einstein equations without destabilizing contributions at the singularities.
We have proven that all admissible solutions result in a fine-tuning of the cosmological constant.

Our main objective has been to show how to stabilize the position of the singularity at parametrically large values compared to the 5D Planck length. We have employed the superpotential method of~\cite{Brandhuber:1999hb,DeWolfe:1999cp} and proposed a family of models that accomplishes this goal. Our stabilizing superpotential allows three types of 4D spectra: continuous, continuous with a mass gap, and discrete. The 4D mass scale $\rho$, controlling the mass gap in the continuous case and the spacing in the discrete one, depends in a double exponential manner on the value of the scalar field at the brane, Eq~(\ref{rhok}), and can thus be naturally suppressed with respect to the 5D mass scale $k$ without fine-tuning. 

Next we have studied in detail the spectra resulting from fluctuations around our family of solutions. As with the case of the background, we have paid close attention to the boundary conditions at the singularity, projecting out solutions that would give contributions at the dynamically generated boundary. We have given analytical forms of the wavefunctions near the brane and the singularity, as well as numerical values for the lowest lying excitations and their profiles.

We have also given a constructive recipe of how to obtain superpotentials that accomplish the stabilization of the hierarchy, a desired spectrum, and an ``end-of-the-world singularity'' that is consistent with the equations of motion. In a first step, one chooses the asymptotic (i.e.~large $\phi$) behavior of $W$. This will determine the asymptotic form of the spectrum. To ensure consistency of the equations of motion, the divergence should be milder than $W\sim e^{2\phi}$. The different asymptotic forms of $W$ and corresponding spectra are summarized~\footnote{A classification of superpotentials giving rise to confining backgrounds was performed in Ref.~\cite{Gursoy:2007cb}.} in Tab.~\ref{default}. In a 
second step, one completes $W$ for smaller values of $\phi$ in such a way as to accomplish a mild hierarchy of the proper distance $y_s$ with respect to the fundamental 5D scale $k$, given by the simple relation Eq.~(\ref{kys2}). Notice that many of the interesting spectra require some kind of exponential behavior at large $\phi$, such that this region does not contribute at all to $k y_s$. At the same time one minimizes the effective 4D potential, Eq.~(\ref{V4}), at the brane at $y=0$ to find the vacuum value of $\phi_0\equiv \phi(0)$.
\begin{table}[tdp]
\begin{center}
\begin{tabular}{| c || c | c | c | c | c |c|}
\hline \vspace{-12pt} & & & & & & \\
 \multirow{2}{*}{$W(\phi)$} & $\leq\phi^2$& $>\phi^2$    & $e^\phi$      &
$e^\phi \phi^\beta$              & $> e^{\phi}\phi^{\frac{1}{2}}$        & $\geq e^{2
\phi}$   \\
 && $<e^\phi$
 &  & {\footnotesize $0 < \beta \le \frac12$}  &
$<e^{2\phi}$
 & 
 \\
\noalign{\hrule height 0.9pt}
$y_s$ & $\infty$  &\multicolumn{5}{c|}{finite}  \\
\hline
$z_s$ &  \multicolumn{4}{c|}{$\infty$} &  \multicolumn{2}{c|}{finite}  \\
\hline
mass & \multicolumn{2}{c|}{\multirow{2}{*}{continuous}} & continuous
& \multicolumn{3}{c|}{discrete} \\
\cline{5-7}
spectrum &  \multicolumn{2}{c|}{} & w/ mass gap  & $m_n \sim
n^{2\beta}$ & \multicolumn{2}{c|}{$m_n \sim n$} \\
\hline
consistent & \multicolumn{5}{c|}{\multirow{2}{*}{yes}} & \multirow{2}{*}{no} \\
solution & \multicolumn{5}{c|}{} &  \\
\hline
\end{tabular}
\end{center}
\caption{\it Spectra resulting from different asymptotic forms of the
superpotential. In the first row we give the asymptotic behavior of
$W(\phi)$, with the strength of the divergence increasing from left to
right ($>$ means ``diverges faster than'', etc). Second and third row
show the finiteness of $y_s$ and $z_s$, with the behavior changing at
$W\sim \phi^2$ and $W\sim e^{\phi}\phi^\frac{1}{2}$ respectively. The
third row shows the spectrum, while in the last one we indicate the
consistency of the solution.}
\label{default}
\end{table}
In a third and final step, one adds a constant of $\mathcal O(k)$ to the superpotential. This adds strong warping near the UV brane, but has no effect whatsoever on the determination of $k y_s$ and $\phi_0$. We have shown that in this way one can warp down the parameter setting the overall scale for the spectrum by a factor $e^{k y_s}$, leading to the desired hierarchy.

There are a number of phenomenological applications which are outside the scope of the present paper but which are worth of future investigations. For the range of the parameter $1<\nu<2$ these applications are common with two brane models, as RS1, but with some peculiarities. In particular graviton (and radion) KK modes are at the TeV scale and they can be produced and decay at LHC by their interaction with matter $\sim h_{\mu\nu}T^{\mu\nu}$, so they are expected to be produced through gluon annihilation~\cite{Lillie:2007yh}. Since there is no IR brane, for soft-wall models to solve the gauge hierarchy problem the Higgs boson (either a scalar doublet or the fifth component of a gauge field in a gauge-Higgs unified model) has to propagate in the bulk and it has to be localized near the singularity for its mass to feel the warping. On the other hand fermions with sizable Yukawa couplings (third generation fermions) should be localized near the singularity as well while first and second generation fermions can propagate at (or near the) UV brane. As we have seen that the first graviton KK mode is localized near the singularity, once produced it is expected to decay into either Higgs or $t\bar t$ pairs. For $\nu=1$ the mass spectrum of fields propagating in the bulk is a continuum above an $\mathcal O(TeV)$ mass gap. This continuum (endowed with a given conformal dimension) can interact with SM fields propagating in the UV brane as operators of a CFT, where the conformal invariance is explicitly broken at a scale given by the mass gap, and can model and describe the unparticle phenomenology. In particular the Higgs embedded into such 5D background can describe the unHiggs theory of Ref.~\cite{Stancato:2008mp} in the presence of a mass gap. In all those cases the strength of electroweak constraints should be an issue. Finally, the case where the spectrum is described by linear Regge trajectories ($\nu=1$, $\beta=1/4$) can give rise to a phenomenological description of AdS/QCD, similar to that of Ref.~\cite{AdS/QCD}, where the QCD scale can be naturally stabilized by the scalar field.

\section*{Acknowledgments}
Work supported in part by the European Commission under the European
Union through the Marie Curie Research and Training Network ``UniverseNet"
(MRTN-CT-2006-035863); by the Spanish Consolider-Ingenio 2010
Programme CPAN (CSD2007-00042); and by CICYT, Spain, under contract
FPA 2008-01430. GG would like to thank IFAE for hospitality during part of this project.
The work of JAC is supported by the Spanish Ministry of Education through a FPU grant.


\end{document}